\documentclass[amsmath,aps,showpacs,a4paper,10pt]{revtex4}

 \usepackage{epsf}
 \usepackage{graphicx}    

\begin{document}

 \newcommand{\be}[1]{\begin{equation}\label{#1}}
 \newcommand{\ee}{\end{equation}}
 \newcommand{\bea}{\begin{eqnarray}}
 \newcommand{\eea}{\end{eqnarray}}
 \def\disp{\displaystyle}

 \def\gsim{ \lower .75ex \hbox{$\sim$} \llap{\raise .27ex \hbox{$>$}} }
 \def\lsim{ \lower .75ex \hbox{$\sim$} \llap{\raise .27ex \hbox{$<$}} }

 \begin{titlepage}

 \begin{flushright}
 arXiv:1005.1445
 \end{flushright}

 \title{\Large \bf Cosmological Constraints on the Modified
 Entropic Force Model}

 \author{Hao~Wei\,}
 \email[\,email address:\ ]{haowei@bit.edu.cn}
 \affiliation{Department of Physics, Beijing Institute
 of Technology, Beijing 100081, China}

 \begin{abstract}\vspace{1cm}
 \centerline{\bf ABSTRACT}\vspace{2mm}
 Very recently, Verlinde considered a theory in which space is
 emergent through a holographic scenario, and proposed that
 gravity can be explained as an entropic force caused by
 changes in the information associated with the positions
 of material bodies. Then, motivated by the Debye model
 in thermodynamics which is very successful in very low
 temperatures, Gao modified the entropic force scenario.
 The modified entropic force (MEF) model is in fact a
 modified gravity model, and the universe can be accelerated
 without dark energy. In the present work, we consider the
 cosmological constraints on the MEF model, and successfully
 constrain the model parameters to a narrow range. We also
 discuss many other issues of the MEF model. In particular,
 we clearly reveal the implicit root to accelerate the
 universe in the MEF model.
 \end{abstract}


 \maketitle

 \end{titlepage}

 \renewcommand{\baselinestretch}{1.1}


\section{Introduction}\label{sec1}

Very recently, Verlinde~\cite{r1} considered a theory in which
 space is emergent through a holographic scenario, and proposed
 that gravity can be explained as an entropic force caused by
 changes in the information associated with the positions of
 material bodies. In this scenario, Verlinde has successfully
 derived the Newton's law of gravitation, the Einstein
 equations, and the law of inertia, from the entropic point of
 view. In fact, the entropic force scenario is similar to the
 old idea of Jacobson~\cite{r2}, but also beyond it in some
 sense. Similar entropic insight into gravity has also been
 made by Padmanabhan~\cite{r3} independently and simultaneously.

Here we briefly mention some key points of the entropic force
 scenario following~\cite{r1}. Motivated by Bekenstein's
 argument~\cite{r4}, Verlinde postulated that the change in
 entropy near the holographic screen is linear in the
 displacement $\Delta x$, namely,
 \be{eq1}
 \Delta S=2\pi k_B\frac{mc}{\hbar}\Delta x\,,
 \ee
 where $m$ is the mass of test particle, whereas $k_B$, $c$
 and $\hbar$ are Boltzmann constant, speed of light and the
 reduced Planck constant, respectively. The effective entropic
 force acting on the test particle due to the change in entropy
 obeys the first law of thermodynamics
 \be{eq2}
 F\Delta x=T\Delta S\,,
 \ee
 where $T$ is the temperature. If one takes the Unruh
 temperature $T$ experienced by an observer in an accelerated
 frame whose acceleration is $a$, i.e.,
 \be{eq3}
 k_B T=\frac{1}{2\pi}\frac{\hbar a}{c}\,,
 \ee
 to be the temperature associated with the bits on the
 holographic screen, from Eqs.~(\ref{eq1})---(\ref{eq3}), it
 is easy to recover the second law of Newton
 \be{eq4}
 F=ma\,.
 \ee
 Considering a sphere as the holographic screen, Verlinde
 assumed that the number of used bits on the holographic screen
 $N$ is proportional to the area $A=4\pi r^2$, i.e.,
 \be{eq5}
 N=\frac{Ac^3}{G\hbar}\,.
 \ee
 According to the equipartition law of energy, the total energy
 inside the screen is
 \be{eq6}
 E=\frac{1}{2}N k_B T\,.
 \ee
 Of course, one can identifies $E$ with the mass $M$ inside the
 screen through
 \be{eq7}
 E=Mc^2.
 \ee
 From Eqs.~(\ref{eq1}), (\ref{eq2}), and (\ref{eq5})---(\ref{eq7}),
 one can recover the Newton's law of gravitation
 \be{eq8}
 F=G\frac{Mm}{r^2}\,,
 \ee
 where $G$ can be identified with the Newton constant now. From
 Eqs.~(\ref{eq3}), (\ref{eq4}) and (\ref{eq8}), it is easy to
 find the gravitational acceleration
 \be{eq9}
 g=\frac{GM}{r^2}\,,
 \ee
 and the temperature
 \be{eq10}
 T=\frac{\hbar}{k_B c}\frac{g}{2\pi}\,.
 \ee
 As shown in~\cite{r1}, a relativistic generalization of the
 presented arguments directly leads to the Einstein equations.
 We strongly refer to the original paper~\cite{r1} for great
 details.

Soon after Verlinde's proposal of entropic force, many relevant
 works appeared. For examples, Cai, Cao and Ohta~\cite{r5},
 Shu and Gong~\cite{r6} derived the Friedmann equations from
 entropic force simultaneously. Smolin~\cite{r7} derived
 the Newtonian gravity in loop quantum gravity. Li and
 Wang~\cite{r8} showed that the holographic dark energy can
 arise in the entropic force scenario. Easson, Frampton and
 Smoot~\cite{r9} considered the entropic accelerating universe
 and the entropic inflation. Tian and Wu~\cite{r10},
 Myung~\cite{r11} discussed the thermodynamics of black holes
 in the entropic force scenario. Vancea and Santos~\cite{r12}
 considered the uncertainty principle from the point of view
 of entropic force. Zhang, Gong and Zhu~\cite{r13},
 Sheykhi~\cite{r14} derived the modified Friedmann equation
 from the corrected entropy. Also, Modesto and
 Randono~\cite{r15} discussed the corrections to Newton's law
 from the corrected entropy. Cai, Liu and Li~\cite{r16}
 considered a unified model of inflation and late-time
 acceleration in the entropic force scenario. For other
 relevant works to entropic force, we refer to
 e.g.~\cite{r17,r18,r19,r36} and references therein.

The works mentioned above are in fact closely following
 Verlinde's proposal of entropic force~\cite{r1}. To be honest,
 here we should also mention the other works which are strongly
 criticizing the entropic force scenario. For instance, the
 author of~\cite{r38} argued that there are some possible flaws
 in Verlinde's idea. In~\cite{r39}, Culetu argued that the
 relativistic Unruh temperature cannot be associated with the
 bits on the screen in the form considered by Verlinde.
 In~\cite{r40}, Hossenfelder argued that some additional
 assumptions made by Verlinde are unnecessary and there are
 some gaps in Verlinde's arguments. In~\cite{r41}, Myung found
 that entropic force does not always imply the Newtonian force
 law, and the connection between Newtonian cosmology and
 entropic force cannot be confirmed. In~\cite{r42}, Li and Pang
 found that inflation is inconsistent with the entropic force
 scenario. In~\cite{r43}, Lee argued that there are some
 inconsistencies in Verlinde's arguments from a classical point
 of view.

So far, we have briefly surveyed the current status of the
 works relevant to the entropic force scenario. It is fair to
 say that the entropic force scenario is still in controversy.
 On the other hand, there is no breakthrough on entropic force
 after Verlinde's proposal~\cite{r1}. A deep insight is needed
 to understand the nature of gravity. In addition, further
 discussions on the entropic force scenario are also desirable.
 Only when more and more results on entropic force are
 available, one can say something conclusively at that time. To
 this end, we would like to contribute our effort and try to
 learn more about the entropic force scenario. In this work, we
 will consider a modified entropic force scenario proposed by
 Gao~\cite{r20}, which has some interesting features. And then,
 we will constrain the modified entropic force scenario with
 the latest observational data.

In~\cite{r20}, Gao noted that statistical thermodynamics
 reveals the equipartition law of energy does not hold in the
 very low temperatures. Instead, as is well known, the Debye
 model~\cite{r21,r22} is very successful in explaining the
 experimental results when the temperatures are very low. Since
 the equipartition law of energy plays an important role in
 the derivation of entropic force, the entropic force
 should be modified for the very weak gravitational fields
 which correspond to very low temperatures. Especially, the
 large-scale universe is in such an extreme weak gravitational
 field, and hence the modified entropic force (MEF) makes
 sense in cosmology.

Following~\cite{r20}, we briefly mention the key points of
 MEF model. Similar to the Debye model~\cite{r21,r22} in
 thermodynamics, one can modify the equipartition law of
 energy in Eq.~(\ref{eq6}) to
 \be{eq11}
 E=\frac{1}{2}N k_B T D(x)\,,
 \ee
 where $D(x)$ is the Debye function which is defined by
 \be{eq12}
 D(x)=\frac{3}{x^3}\int_0^x\frac{y^3}{e^y-1}\,dy\,,
 \ee
 and $x$ is related to the temperature $T$ as
 \be{eq13}
 x\equiv\frac{T_D}{T}\,,
 \ee
 in which $T_D$ is the Debye temperature. By definition, $x$
 is positive. With the modified equipartition law of energy,
 namely Eq.~(\ref{eq11}), similar to the original entropic
 force, one can easily to obtain~\cite{r20}
 \be{eq14}
 g=\frac{GM}{r^2}\frac{1}{D(x)}\,,
 \ee
 in which (nb. Eq.~(\ref{eq10}))
 \be{eq15}
 x=\frac{T_D}{T}=\frac{g_D}{g}\,,
 \ee
 where $g_D\equiv(2\pi\,k_B\,c/\hbar)\,T_D$ is the Debye
 acceleration. Actually, Eq.~(\ref{eq14}) corresponds to the
 modified Newtonian law of gravity. In the limit of strong
 gravitational field, $g\gg g_D$ and hence $x\ll 1$, from
 Eq.~(\ref{eq12}) it is easy to find that $D(x)\to 1$ and the
 Newtonian gravity is recovered. On the other hand, in the
 limit of weak gravitational field, $g\ll g_D$ and hence
 $x\gg 1$, one can see that $D(x)\to\pi^4/(5x^3)$ and then
 $g\propto 1/\sqrt{r}$, which significantly deviates from the
 familiar inverse square law~\cite{r20}. However, as argued
 in~\cite{r20}, one need not to worry about the possibility of
 MEF against the experimental results of the inverse square
 law. Since these experiments testing the inverse square law
 were done on the Earth or in the solar system, which are
 actually in the strong gravitational fields, we have $x\ll 1$
 and $D(x)\to 1$, therefore the deviation from the inverse
 square law are extremely tiny. The significant deviation from
 the inverse square law can only occur in the very large scale
 in the universe where the gravitational fields are very weak,
 and hence it can escape the detection of these experiments testing
 the inverse square law. Of course, this argument relies on a small
 $g_D$. We will justify it later in the present work.

Next, we turn to the cosmological issues in the MEF Model.
 For convenience, we set the units $k_B=c=\hbar=1$ hereafter.
 Using the derivation method in~\cite{r5,r6}, one can find that
 the modified Raychaudhuri equation is given by~\cite{r20}
 \be{eq16}
 4\pi G\left(\rho+p\right)=-\left(\dot{H}-\frac{K}{a^2}\right)
 \left[-2D(x)+\frac{3x}{e^x-1}\right]\,,
 \ee
 where $\rho$ and $p$ are the total energy density and total
 pressure of cosmic fluids, respectively; $K$ is the spatial
 curvature of the universe; $H\equiv\dot{a}/a$ is the Hubble
 parameter; $a=(1+z)^{-1}$ is the scale factor (we have set
 $a_0=1$); $z$ is the redshift; a dot denotes the derivatives
 with respect to cosmic time $t$; the subscript ``0'' indicates
 the present value of the corresponding quantity. Taking into
 account the Hawking temperature $T$ for the universe~\cite{r23}
 \be{eq17}
 T=\frac{H}{\,2\pi}\,,
 \ee
 from Eqs.~(\ref{eq13}) and (\ref{eq15}), it is easy to see
 that
 \be{eq18}
 x=\frac{H_D}{H}\,,
 \ee
 where $H_D=g_D$. On the other hand, the energy conservation
 equation still holds in the MEF model, namely
 \be{eq19}
 \dot{\rho}+3H\left(\rho+p\right)=0\,.
 \ee
 From Eqs.~(\ref{eq16}) and (\ref{eq19}), one can derive the
 corresponding Friedmann equation. It is anticipated
 that Friedmann equation is also modified,
 $H^2\not=(8\pi G\rho)/3$, due to the correction term
 $-2D(x)+3x/(e^x-1)$ in Eq.~(\ref{eq16}). The MEF model is
 in fact a modified gravity model. Gao showed that the MEF
 model can describe the accelerating universe without dark
 energy. We refer to the original paper~\cite{r20} for details.

Since Gao has not considered the constraints on the MEF model
 in~\cite{r20}, we will try to obtain the cosmological
 constraints with the latest observational data in the next
 section. Further, we will discuss some relevant issues of the
 MEF model in Sec.~\ref{sec3}. Finally, we give the brief
 conclusion and some meaningful remarks in Sec.~\ref{sec4}.


 \begin{center}
 \begin{figure}[tbhp]
 \centering
 \includegraphics[width=0.5\textwidth]{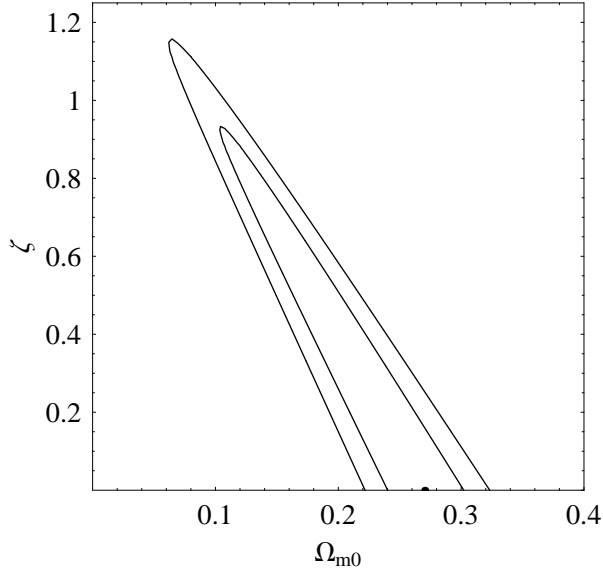}
 \caption{\label{fig1}
 The $68.3\%$ and $95.4\%$ confidence level contours in the
 $\Omega_{m0}-\zeta$ parameter space. The best-fit parameters
 are also indicated by a black solid point. This result is
 obtained by using the data of 557 Union2 SNIa alone.}
 \end{figure}
 \end{center}


\vspace{-10mm} 


\section{Cosmological constraints on the MEF model}\label{sec2}

In this section, we consider the cosmological constraints on
 the MEF model. To this end, we firstly rewrite the equations
 to suitable forms. Notice that we consider a spatially flat
 universe (namely $K=0$) throughout this work.
 From Eqs.~(\ref{eq16}) and (\ref{eq19}), we have
 \be{eq20}
 8\pi G\,d\rho=3\left[-2D(x)+\frac{3x}{e^x-1}\right]dH^2.
 \ee
 As in~\cite{r20}, we consider the universe contains only
 pressureless matter. So, we have
 $\rho=\rho_m=\rho_{m0}\,a^{-3}=\rho_{m0}\,(1+z)^3$. Dividing
 $3H_0^2$ in both sides of Eq.~(\ref{eq20}), we obtain
 \be{eq21}
 \Omega_{m0}\,da^{-3}=\left[-2D(x)+\frac{3x}{e^x-1}\right]dE^2,
 \ee
 where
 \be{eq22}
 \Omega_{m0}\equiv\frac{8\pi G\rho_{m0}}{3H_0^2}\,,~~~~~~~
 E\equiv\frac{H}{H_0}=\frac{\zeta}{x}\,,~~~~~~~
 \zeta\equiv\frac{H_D}{H_0}\,.
 \ee
 Note that $\Omega_{m0}\not=\rho_{m0}/\rho_0=1$, because
 Friedmann equation has been modified in the MEF model, i.e.,
 $H^2\not=8\pi G\rho/3$. By definition, $\zeta$ is positive.
 Finally, we get the differential equation for $E(z)$, namely
 \be{eq23}
 \left[-2D\left(\frac{\zeta}{E}\right)+\frac{3\,\zeta/E}
 {e^{\,\zeta/E}-1}\right]\cdot 2E\,\frac{dE}{dz}
 =3\Omega_{m0}\,(1+z)^2.
 \ee
 In principle, one can numerically find $E(z)$ from this exact
 differential equation, and then fit it to the observational
 data to get the constraints on the MEF model. However, we
 find that it consumes a large amount of time beyond normal
 patience when we scan the grid points in the parameter space,
 mainly due to the hardness of numerically solving the exact
 differential equation~(\ref{eq23}) in which $D(\zeta/E)$ is an
 integral whose upper limit is $\zeta/E$ itself. Therefore, it
 is advisable to find a reliable approximation of the exact
 differential equation~(\ref{eq23}). Note that
 \be{eq24}
 -2D(x)+\frac{3x}{e^x-1}=1-\frac{3}{4}x+{\cal O}\,(x^2),
 \ee
 for any small quantity $x$. On the other hand, in~\cite{r20}
 one might find the hint of a small $\zeta$ (notice that it has
 been chosen to be $10^{-5}$ in~\cite{r20} for example).
 Together with the well-known fact that usually $E(z)$
 increases rapidly when $z$ increases, $\zeta/E$ is a small
 quantity in Eq.~(\ref{eq23}). Therefore, we find that
 an approximation of the exact differential
 equation~(\ref{eq23}) is given by
 \be{eq25}
 \left(1-\frac{3}{4}\frac{\zeta}{E}\right)\cdot 2E\,dE=
 \Omega_{m0}\,da^{-3}.
 \ee
 Integrating Eq.~(\ref{eq25}), we have
 \be{eq26}
 E^2-\frac{3}{2}\,\zeta E=\Omega_{m0}\,a^{-3}+const.,
 \ee
 where $const.$ is the integral constant, which
 can be determined by requiring $E(z=0)=1$. Finally, we find
 that
 \be{eq27}
 E^2-\frac{3}{2}\,\zeta E=\Omega_{m0}\,(1+z)^3
 +\left(1-\frac{3}{2}\,\zeta-\Omega_{m0}\right),
 \ee
 which is a quadratic equation of $E$ in fact. Noting that $E$
 is positive, we solve Eq.~(\ref{eq27}) to get
 \be{eq28}
 E(z)=\frac{3}{4}\,\zeta+\frac{1}{2}\left\{\frac{9}{4}\,\zeta^2
 +4\left[\Omega_{m0}\,(1+z)^3+\left(1-\frac{3}{2}\,\zeta-
 \Omega_{m0}\right)\right]\right\}^{1/2}.
 \ee
 Obviously, when $\zeta\ll 1$, we see that the MEF model reduces to
 the familiar $\Lambda$CDM model in which $E(z)=\left[\Omega_{m0}\,
 (1+z)^3+\left(1-\Omega_{m0}\right)\right]^{1/2}$. Therefore, it is
 not surprising that in~\cite{r20} Gao found the MEF model with
 $\zeta=10^{-5}$ is degenerate to $\Lambda$CDM model. In fact, this
 observation is indeed the key point to understand the reason
 for accelerating the universe without dark energy in the MEF model.

In the following, we consider the cosmological constraints on
 the MEF model from observational data. At first, we use the
 observational data of Type Ia supernovae (SNIa) alone.
 Recently, the Supernova Cosmology Project (SCP) collaboration
 released their Union2 compilation which consists of 557
 SNIa~\cite{r24}. The Union2 compilation is the largest
 published and spectroscopically confirmed SNIa sample to date.
 The data points of the 557 Union2 SNIa compiled in~\cite{r24}
 are given in terms of the distance modulus $\mu_{obs}(z_i)$.
 On the other hand, the theoretical distance modulus is defined as
 \be{eq29}
 \mu_{th}(z_i)\equiv 5\log_{10}D_L(z_i)+\mu_0\,,
 \ee
 where $\mu_0\equiv 42.38-5\log_{10}h$ and $h$ is the Hubble
 constant $H_0$ in units of $100~{\rm km/s/Mpc}$, whereas
 \be{eq30}
 D_L(z)=(1+z)\int_0^z \frac{d\tilde{z}}{E(\tilde{z};{\bf p})}\,,
 \ee
 in which ${\bf p}$ denotes the model parameters. The $\chi^2$
 from 557 Union2 SNIa is given by
 \be{eq31}
 \chi^2_{\mu}({\bf p})=\sum\limits_{i}\frac{\left[
 \mu_{obs}(z_i)-\mu_{th}(z_i)\right]^2}{\sigma^2(z_i)}\,,
 \ee
 where $\sigma$ is the corresponding $1\sigma$ error. The parameter
 $\mu_0$ is a nuisance parameter but it is independent of the data
 points. One can perform an uniform marginalization over $\mu_0$.
 However, there is an alternative way. Following~\cite{r25,r26}, the
 minimization with respect to $\mu_0$ can be made by expanding the
 $\chi^2_{\mu}$ of Eq.~(\ref{eq31}) with respect to $\mu_0$ as
 \be{eq32}
 \chi^2_{\mu}({\bf p})=\tilde{A}-2\mu_0\tilde{B}+\mu_0^2\tilde{C}\,,
 \ee
 where
 $$\tilde{A}({\bf p})=\sum\limits_{i}\frac{\left[\mu_{obs}(z_i)
 -\mu_{th}(z_i;\mu_0=0,{\bf p})\right]^2}
 {\sigma_{\mu_{obs}}^2(z_i)}\,,$$
 $$\tilde{B}({\bf p})=\sum\limits_{i}\frac{\mu_{obs}(z_i)
 -\mu_{th}(z_i;\mu_0=0,{\bf p})}{\sigma_{\mu_{obs}}^2(z_i)}\,,
 ~~~~~~~~~~
 \tilde{C}=\sum\limits_{i}\frac{1}{\sigma_{\mu_{obs}}^2(z_i)}\,.$$
 Eq.~(\ref{eq32}) has a minimum for
 $\mu_0=\tilde{B}/\tilde{C}$ at
 \be{eq33}
 \tilde{\chi}^2_{\mu}({\bf p})=
 \tilde{A}({\bf p})-\frac{\tilde{B}({\bf p})^2}{\tilde{C}}\,.
 \ee
 Since $\chi^2_{\mu,\,min}=\tilde{\chi}^2_{\mu,\,min}$
 obviously, we can instead minimize $\tilde{\chi}^2_{\mu}$
 which is independent of $\mu_0$. The best-fit model
 parameters are determined by minimizing the total $\chi^2$.
 When SNIa is used alone, we have $\chi^2=\tilde{\chi}^2_\mu$
 which is given in Eq.~(\ref{eq33}). As in~\cite{r27,r28},
 the $68.3\%$ confidence level is determined by
 $\Delta\chi^2\equiv\chi^2-\chi^2_{min}\leq 1.0$, $2.3$ and
 $3.53$ for $n_p=1$, $2$ and $3$, respectively, where $n_p$ is
 the number of free model parameters. Similarly, the $95.4\%$
 confidence level is determined by
 $\Delta\chi^2\equiv\chi^2-\chi^2_{min}\leq 4.0$, $6.17$ and
 $8.02$ for $n_p=1$, $2$ and $3$, respectively. In the MEF
 model, there are 2 free model parameters, namely $\Omega_{m0}$
 and $\zeta$. Note that $E(z)$ for the MEF model has been given
 in Eq.~(\ref{eq28}). By minimizing the corresponding $\chi^2$,
 we find the best-fit parameters $\Omega_{m0}=0.2704$ and
 $\zeta=4\times 10^{-7}$, while $\chi^2_{min}=542.683$. In
 Fig.~\ref{fig1}, we present the corresponding $68.3\%$ and
 $95.4\%$ confidence level contours in the $\Omega_{m0}-\zeta$
 parameter space.


 \begin{center}
 \begin{figure}[tbp]
 \centering
 \includegraphics[width=0.5\textwidth]{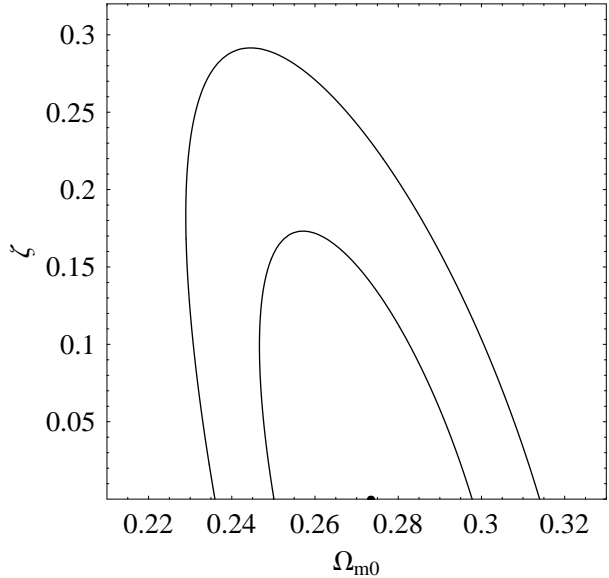}
 \caption{\label{fig2}
 The same as in Fig.~\ref{fig1}, except that this result is
 obtained by using the combined observational data of 557
 Union2 SNIa and the distance parameter $A$ from LSS.}
 \end{figure}
 \end{center}


\vspace{-11mm} 

Next, we add the data from the observation of the large-scale
 structure (LSS). Here we use the distance parameter $A$ of
 the measurement of the baryon acoustic oscillation (BAO)
 peak in the distribution of SDSS luminous red
 galaxies~\cite{r29,r30}, which contains the main information
 of the observations of LSS. The distance parameter $A$ is
 given by
 \be{eq34}
 A\equiv\Omega_{m0}^{1/2}\,E(z_b)^{-1/3}\left[\frac{1}{z_b}
 \int_0^{z_b}\frac{d\tilde{z}}{E(\tilde{z})}\right]^{2/3},
 \ee
 where $z_b=0.35$. In~\cite{r30}, the value of $A$ has been
 determined to be $0.469\,(n_s/0.98)^{-0.35}\pm 0.017$. Here
 the scalar spectral index $n_s$ is taken to be $0.963$, which
 has been updated from the WMAP 7-year (WMAP7) data~\cite{r31}.
 Now, the total $\chi^2=\tilde{\chi}^2_\mu+\chi^2_{LSS}$,
 where $\tilde{\chi}^2_\mu$ is given in Eq.~(\ref{eq33}), and
 $\chi^2_{LSS}=(A-A_{obs})^2/\sigma_A^2$. By minimizing the
 corresponding $\chi^2$, we find the best-fit parameters
 $\Omega_{m0}=0.2733$ and $\zeta=9\times 10^{-8}$, while
 $\chi^2_{min}=542.734$. In Fig.~\ref{fig2}, we present the
 corresponding $68.3\%$ and $95.4\%$ confidence level contours
 in the $\Omega_{m0}-\zeta$ parameter space. Comparing
 Fig.~\ref{fig2} with Fig.~\ref{fig1}, it is easy to see that
 the constraints become much tighter.

\begin{center}
 \begin{figure}[tbhp]
 \centering
 \includegraphics[width=0.5\textwidth]{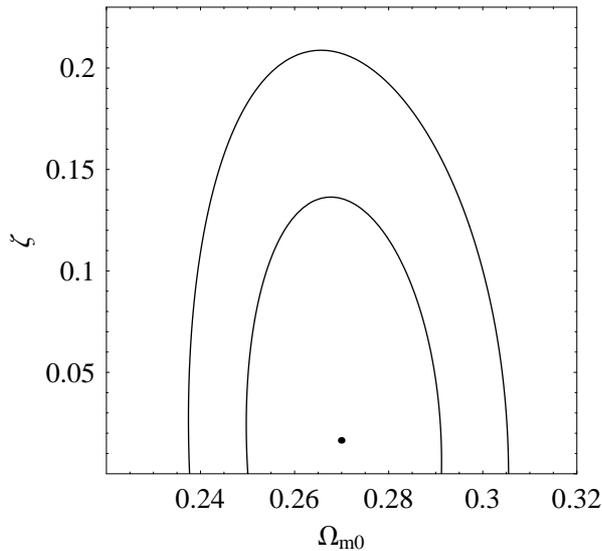}
 \caption{\label{fig3}
 The same as in Fig.~\ref{fig1}, except that this result is
 obtained by using the combined observational data of 557
 Union2 SNIa, the distance parameter $A$ from LSS, and the
 shift parameter $R$ from CMB.}
 \end{figure}
 \end{center}

Then, we further add the data from the observation of the
 cosmic microwave background (CMB). Here we use the the shift
 parameter $R$, which contains the main information of the
 observations of the CMB~\cite{r31,r32,r33}. The shift
 parameter $R$ of the CMB is defined by~\cite{r32,r33}
 \be{eq35}
 R\equiv\Omega_{m0}^{1/2}\int_0^{z_\ast}
 \frac{d\tilde{z}}{E(\tilde{z})}\,,
 \ee
 where the redshift of recombination $z_\ast=1091.3$
 which has been updated in the WMAP7 data~\cite{r31}. The shift
 parameter $R$ relates the angular diameter distance to
 the last scattering surface, the comoving size of the sound
 horizon at $z_\ast$ and the angular scale of the first
 acoustic peak in CMB power spectrum of temperature
 fluctuations~\cite{r32,r33}. The value of $R$ has been updated
 to $1.725\pm 0.018$ from the WMAP7 data~\cite{r31}. Now, the
 total $\chi^2=\tilde{\chi}^2_\mu+\chi^2_{LSS}+\chi^2_{CMB}$,
 where $\chi^2_{CMB}=(R-R_{obs})^2/\sigma_R^2$. By minimizing
 the corresponding $\chi^2$, we find the best-fit parameters
 $\Omega_{m0}=0.2699$ and $\zeta=0.0165$, while
 $\chi^2_{min}=542.879$. In Fig.~\ref{fig3}, we present the
 corresponding $68.3\%$ and $95.4\%$ confidence level contours
 in the $\Omega_{m0}-\zeta$ parameter space. Clearly, the
 constraints become tighter, and the best-fit $\zeta$ significantly
 deviates from zero.


\vspace{-4mm} 

\begin{center}
 \begin{figure}[tbhp]
 \centering
 \includegraphics[width=0.5\textwidth]{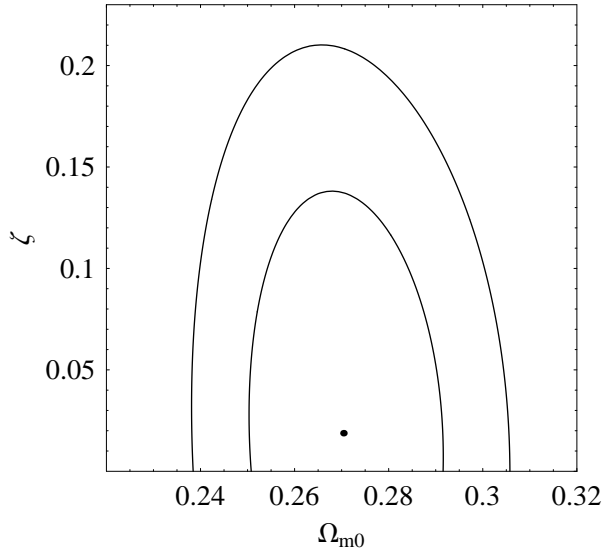}
 \caption{\label{fig4}
 The same as in Fig.~\ref{fig1}, except that this result is
 obtained by using the combined observational data of 557
 Union2 SNIa, the distance parameter $A$ from LSS, the
 shift parameter $R$ from CMB, and 59 Hymnium GRBs.}
 \end{figure}
 \end{center}

Finally, we add the 59 Hymnium Gamma-Ray Bursts
 (GRBs)~\cite{r28}, which can be used to constrain cosmological
 models without the circularity problem. In fact, GRBs are
 a complementary probe to SNIa (see e.g.~\cite{r34} and
 references therein), whose data points are also given in terms
 of the distance modulus $\mu_{obs}(z_i)$, similar to the case
 of SNIa. Therefore, the corresponding $\chi^2_{GRB}$ from
 59 Hymnium GRBs is also given in the same form
 of Eq.~(\ref{eq33}), but the data points are replaced by the
 ones of GRBs. Now, the total
 $\chi^2=\tilde{\chi}^2_\mu+\chi^2_{LSS}+\chi^2_{CMB}+\chi^2_{GRB}$.
 By minimizing the corresponding $\chi^2$, we find the best-fit
 parameters $\Omega_{m0}=0.2704$ and $\zeta=0.0188$, while
 $\chi^2_{min}=566.12$. Note that the number of the total data
 points increased by 59, this $\chi^2_{min}$ is still very good.
 In Fig.~\ref{fig4}, we present the corresponding $68.3\%$ and
 $95.4\%$ confidence level contours in the $\Omega_{m0}-\zeta$
 parameter space. The difference between Figs.~\ref{fig4}
 and~\ref{fig3} is small, mainly due to the relatively
 weak constraint ability of current GRBs sample. This situation
 will be changed when more and more high-quality GRBs are available
 in the future.


\vspace{-10mm} 


\section{Further discussions}\label{sec3}

In the previous section, we have obtained the cosmological
 constraints on the MEF model. Here, we would like to continue
 with further discussions.

\begin{center}
 \begin{figure}[tbhp]
 \centering
 \includegraphics[width=1.0\textwidth]{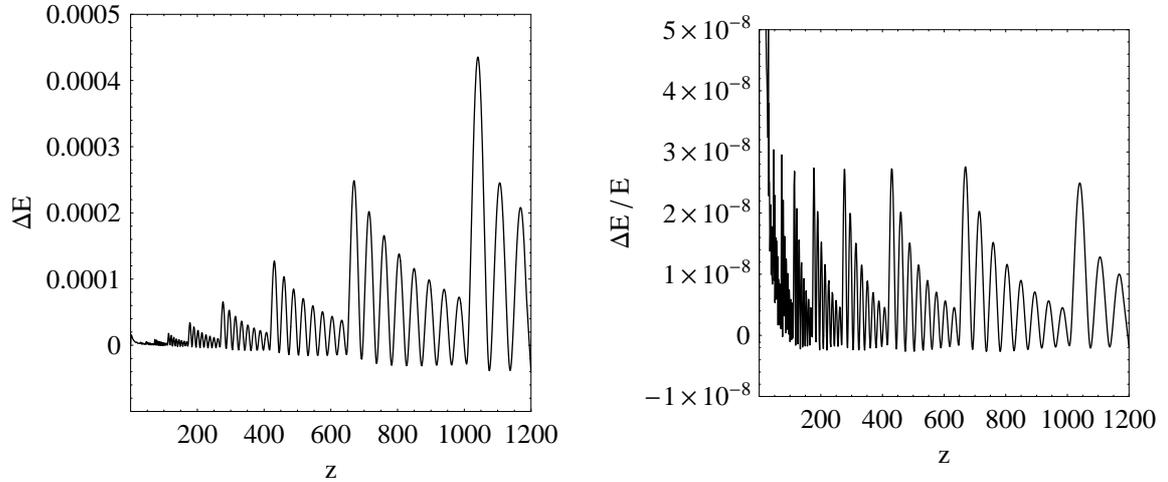}
 \caption{\label{fig5}
 The difference $\Delta E=E_{app}-E_{ex}$ and the relative
 difference $\Delta E/E=(E_{app}-E_{ex})/E_{ex}$ for the case
 with $\zeta=0.017$ and $\Omega_{m0}=0.27$. See the text for
 details.}
 \end{figure}
 \end{center}

\begin{center}
 \begin{figure}[htbp]
 \centering
 \includegraphics[width=1.0\textwidth]{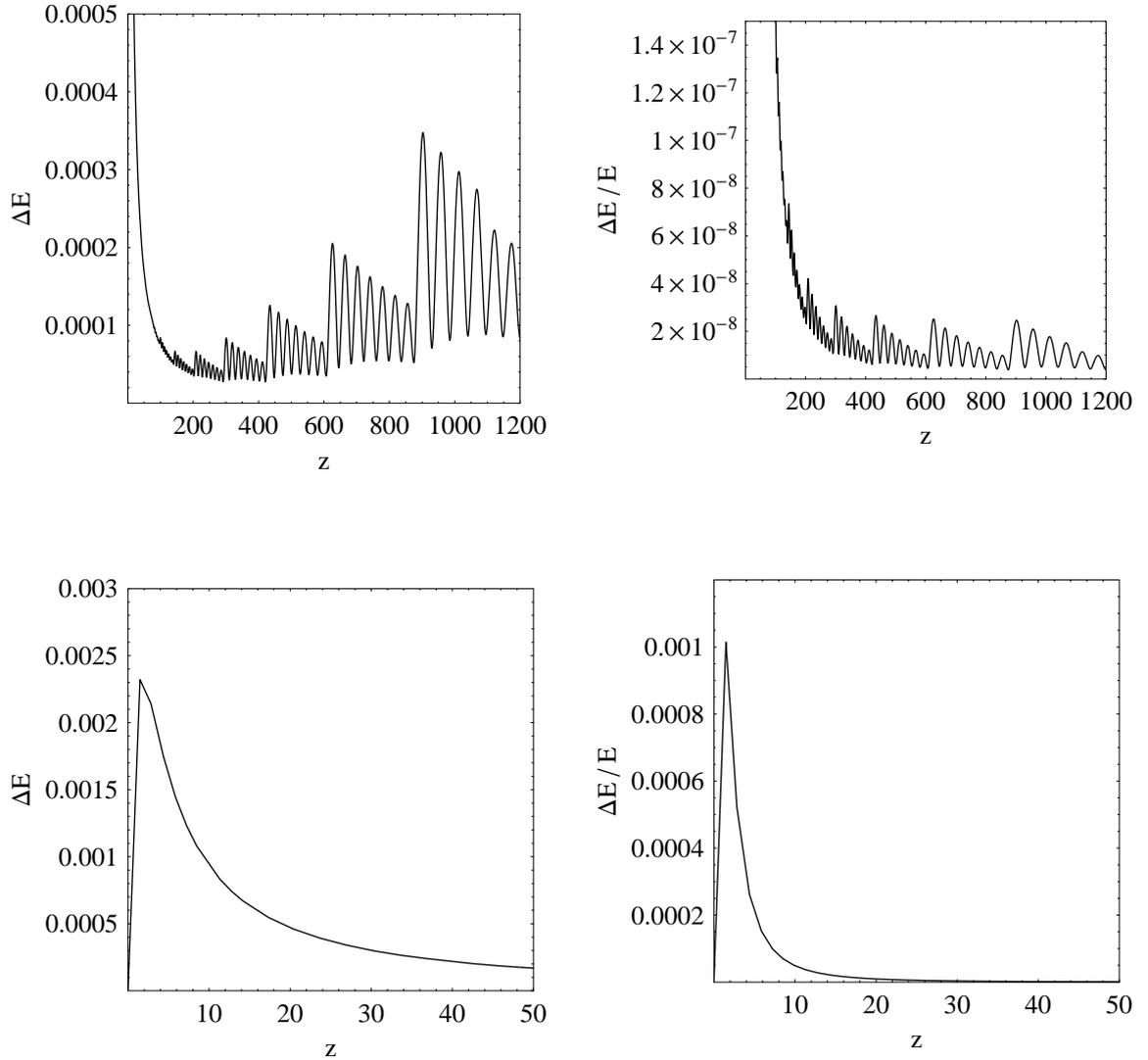}
 \caption{\label{fig6}
 The same as in Fig.~\ref{fig5}, except for the case with
 $\zeta=0.2$ and $\Omega_{m0}=0.27$. The bottom panels are
 the enlarged parts in the redshift range $0\leq z\leq 50$.}
 \end{figure}
 \end{center}

One of the important issues is to justify the approximate
 solution $E(z)$ given in Eq.~(\ref{eq28}). For the very small
 $\zeta$, such as the $\zeta=10^{-5}$ in~\cite{r20} or even smaller,
 we can of course safely use the approximate solution $E(z)$ given
 in Eq.~(\ref{eq28}). The question is when $\zeta$ is not so small,
 can we still safely use the approximate solution $E(z)$ given in
 Eq.~(\ref{eq28})? As the first example, we consider $\zeta=0.017$
 and $\Omega_{m0}=0.27$, which is near to the best fits of the joint
 constraints from SNIa+LSS+CMB and SNIa+LSS+CMB+GRBs. We can
 numerically solve the exact differential equation~(\ref{eq23})
 to find the exact solution of $E(z)$ (of course, the initial
 condition is $E(z=0)=1$). To avoid confusion, we label the exact
 solution from Eq.~(\ref{eq23}) and the approximate solution $E(z)$
 given in Eq.~(\ref{eq28}) as $E_{ex}$ and $E_{app}$, respectively.
 In Fig.~\ref{fig5}, we show the difference
 $\Delta E=E_{app}-E_{ex}$ and the relative difference
 $\Delta E/E=(E_{app}-E_{ex})/E_{ex}$ for the case with
 $\zeta=0.017$ and $\Omega_{m0}=0.27$. Clearly, the difference
 between $E_{app}$ and $E_{ex}$ is very small, and hence we can
 reliably use the approximate solution $E(z)$ given in
 Eq.~(\ref{eq28}). Next, we consider a larger $\zeta$. From
 Figs.~\ref{fig3} and~\ref{fig4}, one can see that the upper edges
 of the $95.4\%$ confidence level contours of the joint constraints
 from SNIa+LSS+CMB and SNIa+LSS+CMB+GRBs extend to $\zeta\sim 0.2$.
 So, we choose the case with $\zeta=0.2$ and $\Omega_{m0}=0.27$
 to be the second example. Again, we present the corresponding
 $\Delta E$ and $\Delta E/E$ in Fig.~\ref{fig6}. One can see
 that the difference between $E_{app}$ and $E_{ex}$ is still fairly
 small, and hence we can also reliably use the approximate solution
 $E(z)$ given in Eq.~(\ref{eq28}). Of course, if the computational
 ability of the computer becomes more powerful, it is undoubtedly
 the best choice to use the exact solution $E_{ex}$ from
 numerically solving Eq.~(\ref{eq23}). When we work with a less
 powerful computer, it is suitable to use the approximate
 solution $E(z)$ given in Eq.~(\ref{eq28}).

\vspace{-5mm} 


\vspace{-10mm} 

\begin{center}
 \begin{figure}[tbp]
 \centering
 \includegraphics[width=1.0\textwidth]{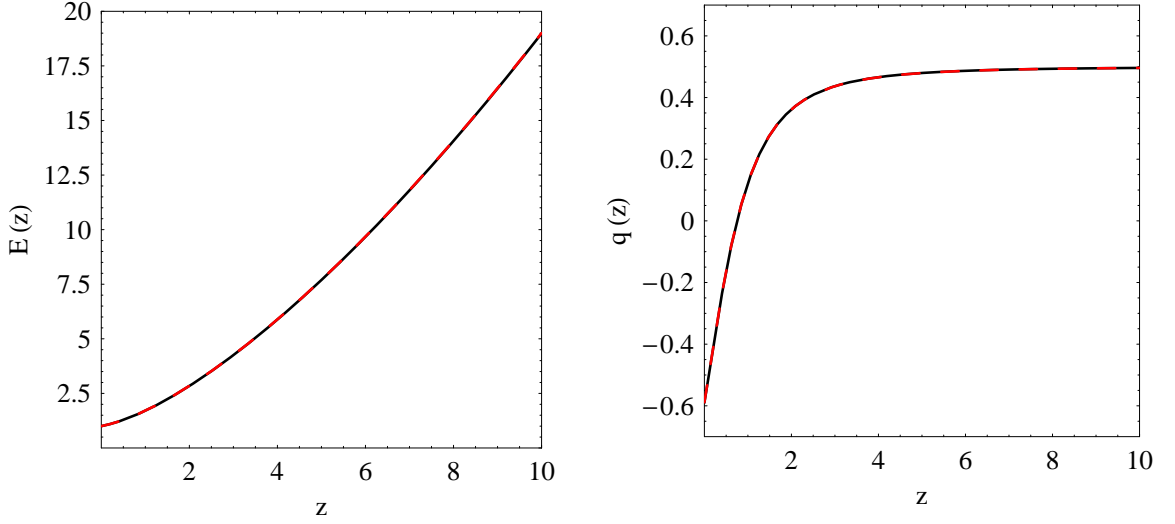}
 \caption{\label{fig7}
 The reduced Hubble parameter $E(z)$ given in Eq.~(\ref{eq28})
 and the corresponding deceleration parameter $q(z)$ for the
 cases with the best-fit parameters of the joint constraints
 from SNIa+LSS+CMB (black solid lines) and SNIa+LSS+CMB+GRBs
 (red dashed lines).}
 \end{figure}
 \end{center}

Next, we turn to the second issue. Although in the text below
 Eq.~(\ref{eq28}) we have shown the key point to understand the
 reason for accelerating the universe without dark energy in
 the MEF model, it is still desirable to show the visualized
 plots. As is well known, the deceleration parameter is
 given by~\cite{r35}
 \be{eq36}
 q\equiv-\frac{\ddot{a}}{aH^2}=-1-\frac{\dot{H}}{H^2}
 =-1+(1+z)\,E^{-1}\frac{dE}{dz}\,.
 \ee
 In Fig.~\ref{fig7}, we plot the reduced Hubble parameter
 $E(z)$ given in Eq.~(\ref{eq28}) and the corresponding
 deceleration parameter $q(z)$ for the cases with the best-fit
 parameters of the joint constraints from SNIa+LSS+CMB (black
 solid lines) and SNIa+LSS+CMB+GRBs (red dashed lines). In
 fact, the plot lines for both cases of SNIa+LSS+CMB and
 SNIa+LSS+CMB+GRBs are heavily overlapped. From
 Fig.~\ref{fig7}, we can clearly see that the deceleration
 parameter $q$ crosses the transition line $q=0$ at redshift
 $z_t=0.75$, and the universe can really be accelerated in the
 late time without dark energy. By the way, we note that the
 usual relation
 $$w_{\rm eff}=-1+\frac{2}{3}(1+z)E^{-1}\frac{dE}{dz}
 =\frac{1}{3}\left(2q-1\right)$$
 does not hold in the MEF model, due to the fact
 that Friedmann equations have been modified. On the contrary,
 $q(z)$ given in Eq.~(\ref{eq36}) holds in any models since
 it comes from definition directly.

Finally, as mentioned in Sec.~\ref{sec1}, the argument that
 MEF can avoid the conflict with the experiments testing the
 inverse square law relies on a small $g_D$. We know that
 $g_D=H_D=\zeta H_0$. In~\cite{r20}, Gao chose a tiny
 $\zeta=10^{-5}$ for example, which can of course make a
 very small $g_D$. However, as shown in this work, the
 allowed $\zeta$ can be in the range $0\leq\zeta\,\lsim\, 0.2$
 within $95.4\%$ confidence level. Notice that the strength of
 gravitational fields are of order $10\ \rm N\cdot kg^{-1}$ on
 the Earth, $10^{-4}\ \rm N\cdot kg^{-1}$ in the solar
 system~\cite{r20}. For the best-fit $\zeta\sim 10^{-2}$, the
 corresponding $g_D=H_D=\zeta H_0\sim 10^{-12}\ \rm N\cdot kg^{-1}$.
 Even for the upper bound $\zeta\sim 10^{-1}$, the corresponding
 $g_D=H_D=\zeta H_0\sim 10^{-11}\ \rm N\cdot kg^{-1}$.
 Therefore, on the Earth or in the solar system, $g\gg g_D$,
 hence $x=g_D/g\ll 1$, and then we have $D(x)\to 1$. Since the
 experiments testing the inverse square law were done on the
 Earth or in the solar system, the deviation from the inverse
 square law are extremely tiny. The significant deviation from
 the inverse square law can only occur in the very large scale
 in the universe where the gravitational fields are very weak,
 and hence it can escape the detection of the experiments
 testing the inverse square law which were done on the Earth
 or in the solar system.


\vspace{-11mm} 


\vspace{-11mm} 


\section{Conclusion and remarks}\label{sec4}

In summary, we considered the cosmological constraints on the
 MEF model in this work, by using the observational data of
 557 Union2 SNIa, the distance parameter $A$ from LSS, the
 shift parameter $R$ from CMB, and 59 Hymnium GRBs. We found
 that the key parameter $\zeta$ in MEF model has been limited
 in a narrow range $0\leq\zeta\,\lsim\, 0.2$ within $95.4\%$
 confidence level. By using the important result given in
 Eq.~(\ref{eq28}), we have clearly shown the key point to
 understand the reason for accelerating the universe without
 dark energy in the MEF model. We showed the MEF model reduces
 to $\Lambda$CDM model when $\zeta\ll 1$. However, the best-fit
 $\zeta$ for the observations SNIa+LSS+CMB and SNIa+LSS+CMB+GRBs
 significantly deviates from zero. This indicates the new
 feature of MEF model different from $\Lambda$CDM model. We
 have justified the approximate solution $E(z)$ given in
 Eq.~(\ref{eq28}). We plotted $E(z)$ and $q(z)$ as
 functions of redshift $z$, and clearly showed that the
 universe can be accelerated in late time without dark energy.
 Finally, we have shown that MEF can avoid the conflict with
 the experiments testing the inverse square law.

After all, some remarks are in order. In the MEF model,
 as shown in this work, the universe can be accelerated without
 dark energy. The only component is dust matter. The MEF model
 is in fact a modified gravity model, similar to the $f(R)$
 models and the braneworld models. The MEF model can be
 degenerate to $\Lambda$CDM model, but it has not an explicit
 cosmological constant in the model. This is a big
 advantage in fact, beyond some $f(R)$ and braneworld models
 in this sense.

Secondly, we point out the possibility to extend the
 original MEF model. In principle, it is not necessary to
 restrict the energy component in the universe to be dust
 matter only. The universe can contain other components, such
 as, dark energy. For example, we can consider a universe
 containing both dust matter and dark energy
 whose equation-of-state parameter (EoS) $w_X$ is a constant.
 In this case, the total energy density
 $\rho=\rho_m+\rho_X=\rho_{m0}\,a^{-3}+\rho_{X0}\,a^{-3(1+w_X)}$.
 Substituting into Eq.~(\ref{eq20}), we have
 \be{eq37}
 \left[-2D\left(\frac{\zeta}{E}\right)+\frac{3\,\zeta/E}
 {e^{\,\zeta/E}-1}\right]\cdot 2E\,\frac{dE}{dz}=
 3\Omega_{m0}\,(1+z)^2+
 3\left(1+w_X\right)\Omega_{X0}(1+z)^{3w_X+2}\,,
 \ee
 where $\Omega_{X0}\equiv(8\pi G\rho_{X0})/(3H_0^2)$. Note that
 $\Omega_{m0}+\Omega_{X0}\not=1$, since Friedmann equation
 has been modified in the MEF model. In principle, one can
 numerically solve Eq.~(\ref{eq37}) to get $E(z)$.
 Similar to the original MEF model, we find that
 an approximation of the exact differential
 equation~(\ref{eq37}) is given by
 \be{eq38}
 \left(1-\frac{3}{4}\frac{\zeta}{E}\right)\cdot 2E\,dE=
 \Omega_{m0}\,da^{-3}+\Omega_{X0}\,da^{-3(1+w_X)}\,.
 \ee
 Integrating Eq.~(\ref{eq38}), we have
 \be{eq39}
 E^2-\frac{3}{2}\,\zeta E=\Omega_{m0}\,a^{-3}
 +\Omega_{X0}\,a^{-3(1+w_X)}+const.,
 \ee
 where $const.$ is the integral constant, which
 can be determined by requiring $E(z=0)=1$. Finally, we find
 that
 \be{eq40}
 E^2-\frac{3}{2}\,\zeta E=\Omega_{m0}\,(1+z)^3+
 \Omega_{X0}\,(1+z)^{3(1+w_X)}
 +\left(1-\frac{3}{2}\,\zeta-\Omega_{m0}-\Omega_{X0}\right),
 \ee
 which is a quadratic equation of $E$ in fact. Noting that $E$
 is positive, we solve Eq.~(\ref{eq40}) to get
 \be{eq41}
 E(z)=\frac{3}{4}\,\zeta+\frac{1}{2}\left\{\frac{9}{4}\,\zeta^2
 +4\left[\Omega_{m0}\,(1+z)^3+\Omega_{X0}\,(1+z)^{3(1+w_X)}
 +\left(1-\frac{3}{2}\,\zeta-\Omega_{m0}-\Omega_{X0}\right)
 \right]\right\}^{1/2}.
 \ee
 In fact, this is just a simple example. One can include any
 type of dark energy, for instance, the CPL dark energy whose
 EoS is given by $w_{de}=w_0+w_a(1-a)$, quintessence, phantom,
 k-essence, hessence, (generalized) Chaplygin gas,
 holographic/agegraphic dark energy, vector-like dark energy,
 spinor dark energy, and so on. Therefore, we would like to
 give the more general formulae. In this case, the total
 energy density
 $\rho=\rho_m+\rho_{de}=\rho_{m0}\,a^{-3}+\rho_{de,0}\,f(a)$,
 where $f(a)$ can be any function of $a$ which satisfies
 $f(a=1)=1$. Substituting into Eq.~(\ref{eq20}), we obtain
 \be{eq42}
 \left[-2D\left(\frac{\zeta}{E}\right)+\frac{3\,\zeta/E}
 {e^{\,\zeta/E}-1}\right]\cdot 2E\,\frac{dE}{dz}=
 3\Omega_{m0}\,(1+z)^2-
 \Omega_{de,0}\,(1+z)^{-2}f^\prime\,,
 \ee
 where $f^\prime\equiv df/da$, and
 $\Omega_{de,0}\equiv(8\pi G\rho_{de,0})/(3H_0^2)$. Note
 again that $\Omega_{m0}+\Omega_{de,0}\not=1$, since Friedmann
 equation has been modified in the MEF model. Eq.~(\ref{eq42})
 is the exact differential equation, which can be used to
 find the exact $E(z)$ numerically. Also, we give
 the corresponding approximate solution as
 \be{eq43}
 E(z)=\frac{3}{4}\,\zeta+\frac{1}{2}\left\{\frac{9}{4}\,\zeta^2
 +4\left[\Omega_{m0}\,(1+z)^3+
 \Omega_{de,0}\,f\left(\frac{1}{1+z}\right)
 +\left(1-\frac{3}{2}\,\zeta-\Omega_{m0}-\Omega_{de,0}\right)
 \right]\right\}^{1/2}.
 \ee
 Similarly, if one need to add other component, such as
 radiation, it is not a hard work. In fact, we can give the
 most general formulae. In this case, the total energy density
 $\rho=\rho_0\,f(a)$, where $f(a)$ can be any function of $a$
 which satisfies $f(a=1)=1$. Substituting into Eq.~(\ref{eq20}),
 we have
 \be{eq44}
 \left[-2D\left(\frac{\zeta}{E}\right)+\frac{3\,\zeta/E}
 {e^{\,\zeta/E}-1}\right]\cdot 2E\,\frac{dE}{dz}=
 -\Omega_0\,(1+z)^{-2}f^\prime\,,
 \ee
 where $\Omega_0\equiv(8\pi G\rho_0)/(3H_0^2)$. Note again that
 $\Omega_0\not=1$, since Friedmann equation has been modified
 in the MEF model, namely $H^2\not=(8\pi G\rho)/3$.
 Eq.~(\ref{eq44}) is the exact differential equation, which can
 be used to find the exact $E(z)$ numerically. Also, we give
 the corresponding approximate solution as
 \be{eq45}
 E(z)=\frac{3}{4}\,\zeta+\frac{1}{2}\left\{\frac{9}{4}\,\zeta^2
 +4\left[\,\Omega_0\,f\left(\frac{1}{1+z}\right)
 +\left(1-\frac{3}{2}\,\zeta-\Omega_0\right)\right]\right\}^{1/2}.
 \ee
 In fact, noting that $E=H/H_0$, Eq.~(\ref{eq45}) can be
 regarded as the approximate modified Friedmann equation in
 the MEF model. If $\zeta\ll 1$, Eq.~(\ref{eq45}) reduces to
 \be{eq46}
 H^2=\frac{8\pi G}{3}\,\rho+\Lambda_{\rm eff}\,,
 \ee
 where $\Lambda_{\rm eff}=\left(1-\Omega_0\right)H_0^2=const.$
 is actually an effective cosmological constant. Therefore, in
 the most general case, we can clearly reveal the implicit root
 to accelerate the universe in the MEF model, regardless of the
 energy components in the universe. An effective cosmological
 constant is the intrinsic feature (we refer to e.g.~\cite{r23}
 for a previous insight). The other exotic features of the MEF
 model could emerge only when $\zeta$ significantly deviates
 from zero.

Thirdly, we would like to say some words on the understanding
 of the original entropic force model~\cite{r1} and the
 modified entropic force model~\cite{r20}. In fact, entropic
 force is just a new perspective to gravity, from the
 thermodynamical point of view. So, the original entropic force
 can only recover all results of the usual (Newton and Einstein)
 gravity. The only new thing is the reversed logic which might
 reveal the nature of gravity. In the original entropic force
 model~\cite{r1}, using the fundamental assumptions
 Eqs.~(\ref{eq1}), (\ref{eq3}), (\ref{eq5}) and~(\ref{eq6}),
 Verlinde derived the Newton's law of gravitation
 Eq.~(\ref{eq8}) for the (non-relativistic) Euclidean spacetime
 in section~3 of~\cite{r1}, and also derived the Einstein
 gravitational equations for any (relativistic) curved
 spacetime in section~5 of~\cite{r1}. On the other hand, the
 Friedmann equations were derived in~\cite{r5,r6} for the
 Friedmann-Robertson-Walker (FRW) universe. There is {\em no}
 any mixing here. We should mention that both the original
 entropic force~\cite{r1} and the modified
 entropic force~\cite{r20} cannot be understood in {\em only}
 non-relativistic or relativistic cases. In fact, they
 are equivalent to gravity itself in all cases. As the
 usual understanding, the Newton's law of gravitation is just
 the approximation of Einstein gravitational equations in the
 (non-relativistic) small scale limit, whereas the Friedmann
 equations are just the special case of Einstein gravitational
 equations in the cosmic scale (homogeneous and isotropic
 spacetime). The situation is similar in the modified entropic
 force model. Using the fundamental assumptions
 Eqs.~(\ref{eq1}), (\ref{eq3}), (\ref{eq5}) and~(\ref{eq11}),
 in~\cite{r20} Gao derived the Newton's law of gravitation
 Eq.~(\ref{eq14}) for the (non-relativistic) Euclidean
 spacetime, and also derived the second Friedmann
 equation~(\ref{eq16}) for the FRW universe. Note that the
 first Friedmann equation can be derived from the second
 Friedmann equation~(\ref{eq16}) and the energy conservation
 equation~(\ref{eq19}). On the other hand, following Verlinde's
 derivations in section~5 of~\cite{r1}, one can derive the
 corresponding (modified) Einstein gravitational equations for
 any (relativistic) curved spacetime. In fact, this is just
 the lacked sector in the modified entropic force model.
 However, it is available in principle, although it has
 not been given in the literature. In the modified entropic
 force model, there is {\em no} any mixing too. The modified
 Newton's law of gravitation Eq.~(\ref{eq14}) is just the
 approximation of the (lacked but available in principle)
 modified Einstein gravitational equations in the
 (non-relativistic) small scale limit, whereas the modified
 second Friedmann equation~(\ref{eq16}) is just the special
 case of the (lacked but available in principle) modified
 Einstein gravitational equations in the cosmic scale
 (homogeneous and isotropic spacetime).

Fourthly, we said that the MEF model is similar to
 $f(R)$-gravity or braneworld scenario. Notice that they are
 similar {\em only} in the sense that the gravity has been
 modified in these models. Of course, both $f(R)$-gravity and
 braneworld scenario were derived from the known actions,
 whereas the action for MEF is still lacked in the literature.
 However, as mentioned above, following Verlinde's derivations
 in section~5 of~\cite{r1}, in principle one can derive the
 corresponding (modified) Einstein gravitational equations for
 any (relativistic) curved spacetime. Once this lacked sector
 has been done, the explicit action is ready. Since the
 present work focuses on cosmology in the MEF model, we leave
 this task to future works.

Fifthly, as mentioned in this work, in the MEF model, there is
 {\em no} dark energy in fact. The universe is matter-dominated
 always. The expansion of our universe is accelerated due to
 the fact that gravity has been modified. In the
 non-relativistic case, there is no dark energy too, but
 gravity is also modified. However, as mentioned in this work,
 this modification to Newtonian gravity is negligible on the
 Earth or in the solar system. In the larger scale, the
 modified gravity is described by Eq.~(\ref{eq14}). As shown
 in~\cite{r37}, the Debye entropic force can be an alternative
 to the modified Newtonian dynamics (MOND) to explain the
 rotational velocity curves of spiral galaxies. In fact,
 the MEF model~\cite{r20} and the Debye entropic force
 model~\cite{r37} are very similar. So, it is anticipated that
 the ``non-relativistic cosmology'' of the MEF model could be
 an alternative to dark matter, which is usually invoked to
 explain the rotational velocity curves of spiral galaxies.

Finally, we admit that the entropic force proposed by Verlinde
 is based on several unproved hypotheses, and it is still
 controversial in the physical community. On the other hand,
 the Debye model in the thermodynamics has not been used in
 the gravity theory previously. However, in the history, many
 great theories also appeared controversially in their
 beginning. Therefore, we consider that it is better to keep
 an open mind to these speculative attempts.


\section*{ACKNOWLEDGEMENTS}
We thank the anonymous referee for quite useful comments
 and suggestions, which help us to improve this work. We are
 grateful to Professors Rong-Gen~Cai, Shuang~Nan~Zhang and
 Miao~Li for helpful discussions. We also thank Minzi~Feng,
 as well as Xiao-Peng~Ma and Bo~Tang, for kind help and
 discussions. This work was supported in part by NSFC under
 Grant No.~10905005, the Excellent Young Scholars Research
 Fund of Beijing Institute of Technology, and the Fundamental
 Research Fund of Beijing Institute of Technology.

\renewcommand{\baselinestretch}{1.1}


\end{document}